\documentclass[10pt,journal,twocolumn,final]{IEEEtran}
\usepackage{xspace,amsthm,amsmath,amsfonts,syntonly}
\usepackage{amssymb}
\usepackage{epsfig} 
\usepackage{epstopdf}
\usepackage{epsf,graphics,graphicx}
\usepackage{subcaption}
\usepackage{cite,bm,color,url,textcomp}
\usepackage{float}
\usepackage{algorithmicx,algorithm,algpseudocode}
\usepackage{url}
\include{com}

\begin{document}

\title{Online Learning of Trellis Diagram Using Neural Network for Robust Detection and Decoding}

\author{Jie Yang, \textit{Student Member, IEEE,}  \quad Qinghe Du, \textit{Student Member, IEEE,} \quad Yi Jiang, \textit{Member, IEEE,}\\
}
\maketitle

\footnote{The work was supported by National Natural Science Foundation of China Grant No. 61771005. Partial material of this paper was published in IEEE/CIC International Conference on Communications (ICCC) 2021, Xiamen, China. \emph{(Corresponding author: Yi Jiang)}

J. Yang, Q. Du, and Y. Jiang are with Key Laboratory for Information Science of Electromagnetic Waves (MoE), School of Information Science and Technology, Fudan University, Shanghai, China (E-mails: jieyang19@fudan.edu.cn, qhdu20@fudan.edu.cn, yijiang@fudan.edu.cn).}
\begin{abstract}

This paper studies machine learning-assisted maximum likelihood (ML) and maximum a posteriori (MAP) receivers for a communication system with memory, which can be modelled by a trellis diagram. The prerequisite of the ML/MAP receiver is to obtain the likelihood of the received samples under different state transitions of the trellis diagram, which relies on the channel state information (CSI) and the distribution of the channel noise. We propose to learn the trellis diagram real-time using an artificial neural network (ANN) trained by a pilot sequence. This approach, termed as the online learning of trellis diagram (OLTD), requires neither the CSI nor statistics of the noise, and can be incorporated into the classic Viterbi and the BCJR algorithm. 
It is shown to significantly outperform the model-based methods in non-Gaussian channels. It requires much less training overhead than the state-of-the-art methods, and hence is more feasible for real implementations. As an illustrative example, the OLTD-based BCJR is applied to a Bluetooth low energy (BLE) receiver trained only by a 256-sample pilot sequence. Moreover, the OLTD-based BCJR can accommodate for turbo equalization, while the state-of-the-art BCJRNet/ViterbiNet cannot. As an interesting by-product, we propose an enhancement to the BLE standard by introducing a bit interleaver to its physical layer; the resultant improvement of the receiver sensitivity can make it a better
fit for some Internet of Things (IoT) communications.
\end{abstract}


\IEEEpeerreviewmaketitle

\section{Introduction}\label{introduction}

Reliable detection and decoding is essential for any communication systems. For a single-carrier system in an inter-symbol interference (ISI) channel, which can be modeled by a finite-state trellis diagram, the classic design of a maximum likelihood (ML) receiver relies on the likelihood function of the state transitions in the trellis diagram \cite{john2008digital}. In this paper, we propose to learn the likelihood function using an artificial neural network (ANN) based on a pilot sequence of moderate length, requiring neither the  channel state information (CSI) nor the statistics of the noise.

As related works, machine learning-assisted wireless communications have attracted broad attentions in recent years\cite{o'shea2017an, mao2018deep, gunduz2019machine, tran2019federated,park2020meta}, such as machine learning-assisted channel decoding \cite{7852251,8437530,8259241,nachmani2018deep, gruber2017on} and symbol detection in a multi-input multi-output (MIMO) system \cite{samuel2017deep, samuel2019learning,8934725,o'shea2017deep}. Deep learning algorithms are found to be more effective in addressing the difficult problem of symbol detection with incomplete CSI \cite{liao2019deep,farsad2018neural,aoudia2018end,ye2020deep}. The aforementioned works, however, attempt to substitute a whole communication system by an ANN, which requires a large amount of training data (a lengthy pilot sequences of tens of thousand samples or more), way too much to be practically feasible. The notion of the model-driven method, which combines machine learning techniques and model-based expert knowledge, is introduced in \cite{8715338} to better incorporate machine learning techniques into a communication system. The efficacy of this type of model-driven methods is demonstrated in \cite{8509622, 9075976}. 

The recent works by Shlezinger {\it et. al.}  \cite{shlezinger2020data, shlezinger2020viterbinet} advocate to use a (relatively simple) neural network to substitute only the channel-dependent part of the Viterbi \cite{forney1973the}  and the BCJR receiver \cite{bahl1974optimal}. The resultant algorithms, the so-termed ViterbiNet and  BCJRNet, train a neural network to learn the {\em a posteriori} probability (APP)  of the state transitions given the received samples and use the finite mixture model (FMM) \cite{mclachlan2000finite} to estimate the marginal probability density of the channel output, assuming that the channel noise is Gaussian. Therefore, the ViterbiNet and BCJRNet require only several thousand training samples, which are significantly less than that in \cite{liao2019deep,farsad2018neural}, but may still be too much to be practically competitive compared with the conventional model-based method.


In this paper, we first consider the same problem as addressed in  \cite{shlezinger2020data, shlezinger2020viterbinet}. We adopt the notion of integrating a simple neural network into a communication system as advocated by Shlezinger {\it et. al.} and propose a new method, termed as the online learning of trellis diagram (OLTD), that can also be integrated into the Viterbi algorithm  \cite{forney1973the} and the BCJR algorithm \cite{bahl1974optimal}. The resultant OLTD-based Viterbi and OLTD-based BCJR differ from the ViterbiNet and BCJRNet in that the ANN is used to learn the likelihoods of the received samples under different state transitions, rather than the APPs. Therefore, we do not need to assume the channel noise to be  Gaussian nor to estimate the marginal distribution of the channel output; thus, our proposed method is simpler, robuster, and requires a substantially shorter pilot sequence.



To show the practical feasibility of  the proposed OLTD method, we apply it to the physical layer (PHY) of the Bluetooth Low Energy (BLE)  protocol\cite{au2019bluetooth}. A BLE system adopts the coded Gaussian frequency shift keying (GFSK) modulation, which belongs to the family of continuous phase modulation (CPM) \cite{sundberg1986continuous}. By modeling the GFSK modulation process with a trellis diagram, we employ the OLTD method to learn the likelihoods of the state transition associated with each received sample based on the 256-sample pilot sequence as regulated in the BLE protocol\cite[Ch. 6, PartB]{Bluetooth5}, and then use the conventional Viterbi or BCJR 
algorithm to recover the information bits.  This study shows that our proposed neural network assisted receiver can work for a real wireless protocol. 

We further introduce a bit-interleaver to the coded GFSK system, for which we combine seamlessly the OLTD method with turbo equalization  \cite{okada2002turbo-equalization, wang2001turbo, okada2001turbo} to achieve significantly improved receiver sensitivity than the conventional Bluetooth. Unlike the previous neural network based methods that unfold each iteration with one layer of the neural network \cite{8259241, gruber2017on}, the OLTD-based method obtains the likelihood of each connected branch in the trellis diagram only once through the whole iterative process. In contrast, the BCJRNet \cite{shlezinger2020viterbinet} assumes that all the coded bits have equal probability and therefore is not suitable for turbo equalization as explained in Section \ref{IV-III}.

The contributions of this paper are summarized as follows:
\begin{itemize}
\item We show that what a neural network needs to learn about the trellis diagram is {\it the normalized likelihoods } of the received sample, 
rather than the {\it a posteriori} probabilities of state transitions as adopted in the BCJRNet/ViterbiNet \cite{shlezinger2020data, shlezinger2020viterbinet}. Owing to this insight, our proposed OLTD method is computationally simpler, robuster, and more versatile than the state-of-the-art methods. We also show that adopting an  ANN with only one hidden-layer is sufficient for the OLTD. 

\item In contrast to the BCJRNet and ViterbiNet \cite{shlezinger2020data, shlezinger2020viterbinet}, our proposed OLTD does not need to compute the marginal distribution of received samples, nor does it assumes the statistics of the channel noise or the {\it a priori} probabilities of the transmitted symbols; thus, the OLTD is computationally much simpler and robuster.

\item The OLTD-based BCJR method can be seamlessly incorporated into a turbo equalizer for significantly improved performance. But the BCJRNet method cannot, as simulated and analyzed in Section \ref{IV-III}. In this sense, our proposed method is more versatile.

\item Based on a  pilot sequence of a practical length, our neural network-based method can outperform the model-based approaches in channels with non-Gaussian noise or interference, as illustrated by the numerical simulations.

\item As an interesting by-product, this study indicates a possible enhancement to the BLE standard, i.e., to introduce a bit interleaver between the convolutional encoder and the GFSK modulator for much improved reliability, which can make BLE a better candidate for some Internet of Things (IoT) communications \cite{atzori2010the}. 
\end{itemize}

The remainder of this paper is organized as follows: Section \ref{System-Model} introduces the system model and briefly review the BCJR algorithm and Viterbi algorithm. Section \ref{sec:OLTD-based-BCJR} explains how to train the OLTD and how the OLTD-based BCJR/Viterbi algorithm works online. Section \ref{IV} introduces the OLTD-based turbo equalization and its application to a bit-interleaved coded GFSK system. In Section \ref{V}, the simulation results are given to verify the effectiveness of OLTD-based method and show the superior performance of the proposed OLTD-based BCJR/Viterbi and OLTD-based turbo algorithm. The conclusion is given in Section \ref{conclusion}.

\section{System Model and Preliminaries} \label{System-Model}
\subsection{An ISI Channel Model}\label{signal-Model}
The received signal in an ISI channel can be expressed as
\begin{equation}
y_k = \sum_{l=0}^{L-1}h_l x_{k-l}+z_k,k=1,2,\dots,N,
\label{eq1}
\end{equation}
where $L$ is the channel length, $h_l, 0 = 1,\cdots,L-1$ are the channel coefficients, and $z_k$ denotes the i.i.d additive noise, which is not necessarily Gaussian.

\begin{figure}[!htb]
\centering
\hspace{0em}
\begin{tabular}{l}
{\psfig{figure=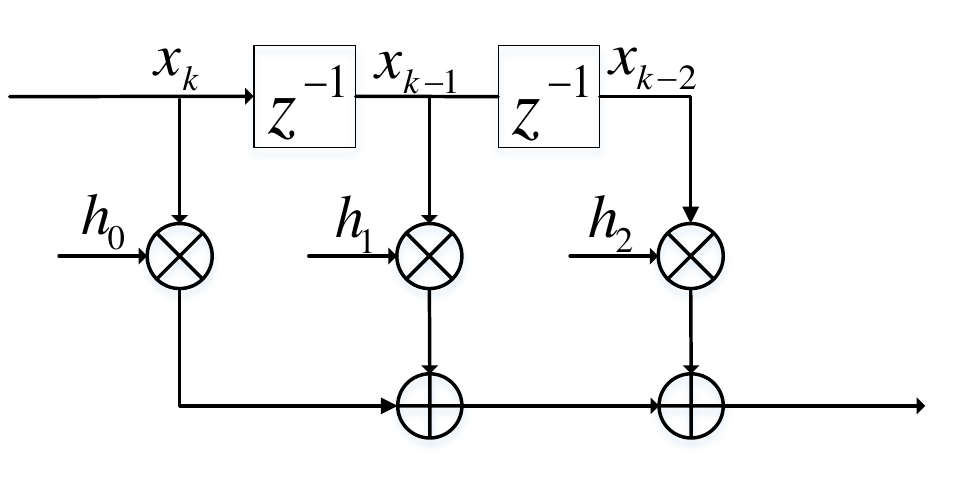 ,width=2.3in}}
\end{tabular}
\caption{The tapped delay line model for an ISI channel with length $L=3$.}
\label{Fig1}
\end{figure}

The ISI channel can be modeled as a tapped delay line as shown in Fig. \ref{Fig1}. Owing to the shift register structure, the channel can be modelled by a trellis diagram \cite{john2008digital}.

\begin{figure}[!ht]
\centering
\hspace{0em}
\begin{tabular}{l}
{\psfig{figure=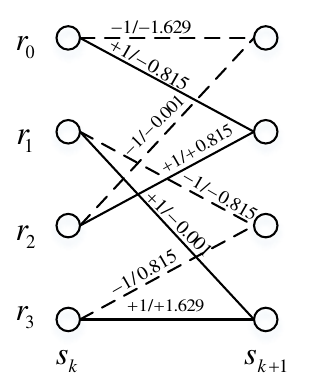, width=1.8in}}
\end{tabular}
\caption{The trellis diagram of the ISI channel with length $L = 3$ and the BPSK input; the solid line represents $u_k = 1$, the dashed line represents $u_k = 0$. }
\label{Fig2}
\end{figure}

As the transmitted symbols are drawn from a set of ${\mathcal X}$, the ISI channel modeled in (\ref{eq1}) can be represented by a trellis diagram consisting of a set ${\mathcal S}$  of the states  with cardinality  $|\mathcal{S}| = |{\mathcal X}|^{L-1}$. Denote the state set at time $k$ as $s_k$, which corresponds to the combination of $L-1$ symbols $x_{k-L+1},...,x_{k-1}$. The state transitions\ $s_k \rightarrow s_{k+1}$ is associated with the output signal $v_k = \sum_{l=0}^{L-1} h_l x_{k-l}$. As an illustrative example, consider an ISI channel with coefficients $h_0 = 0.407$, $h_1= 0.815$, and $h_2 = 0.407$ and the binary phase shift key (BPSK) signal as the input. The trellis consists of $|\mathcal{S}| = 4$ states, with $r_0 = (-1,-1)$, $r_1 = (+1,-1)$, $r_2 = (-1,+1)$, $r_3= (+1,+1)$, as shown in Fig. \ref{Fig2}. The numbers on the branches represent $u_k$ and $v_k$. Take the state $r_0$ for example: if the input symbol $u_k = -1$, the branch output $v_k = -h_0-h_1-h_2=-1.629$; if  $u_k = +1$, $v_k = h_0-h_1-h_2=-0.815$.

\subsection{The GFSK Modulation} \label{coded-GFSK}
The trellis diagram can also be used to model a modulation with memory. As an example, we review the GFSK, which is used in the PHY of Bluetooth \cite{au2019bluetooth}.
The GFSK with bandwidth $1/T$ sampled at $t=nT$ is
\begin{equation} \label{eq16}
x_n = e^{j\phi_n},n=1,2\cdots, N,
\end{equation}
where
\begin{equation}
\phi_n = \pi \mathsf{h} \sum_{m=1}^{n-1} I_m+2\pi \mathsf{h} I_nq(T). \label{eqphin}
\end{equation}
Here $I_n \in \{0,1 \}$ is the information bits, ${\sf h}$ is the modulation index, and the pulse shaping function
\begin{equation}
q(t)=\frac{\int_{0}^tQ\left( 2\pi B\frac{\tau-\frac{T}{2}}{\sqrt{\ln 2}}\right)-Q\left(2\pi B\frac{\tau+\frac{T}{2}}{\sqrt{\ln 2}}\right)d\tau}{2T}
\end{equation}
with $Q(x)=\frac{1}{\sqrt{2\pi}}\int_{x}^\infty e^{\frac{-\tau^2}{2}}d\tau$. 

\begin{figure}[!ht]
\hspace{0em}
\centering
\begin{tabular}{c}
\psfig{figure=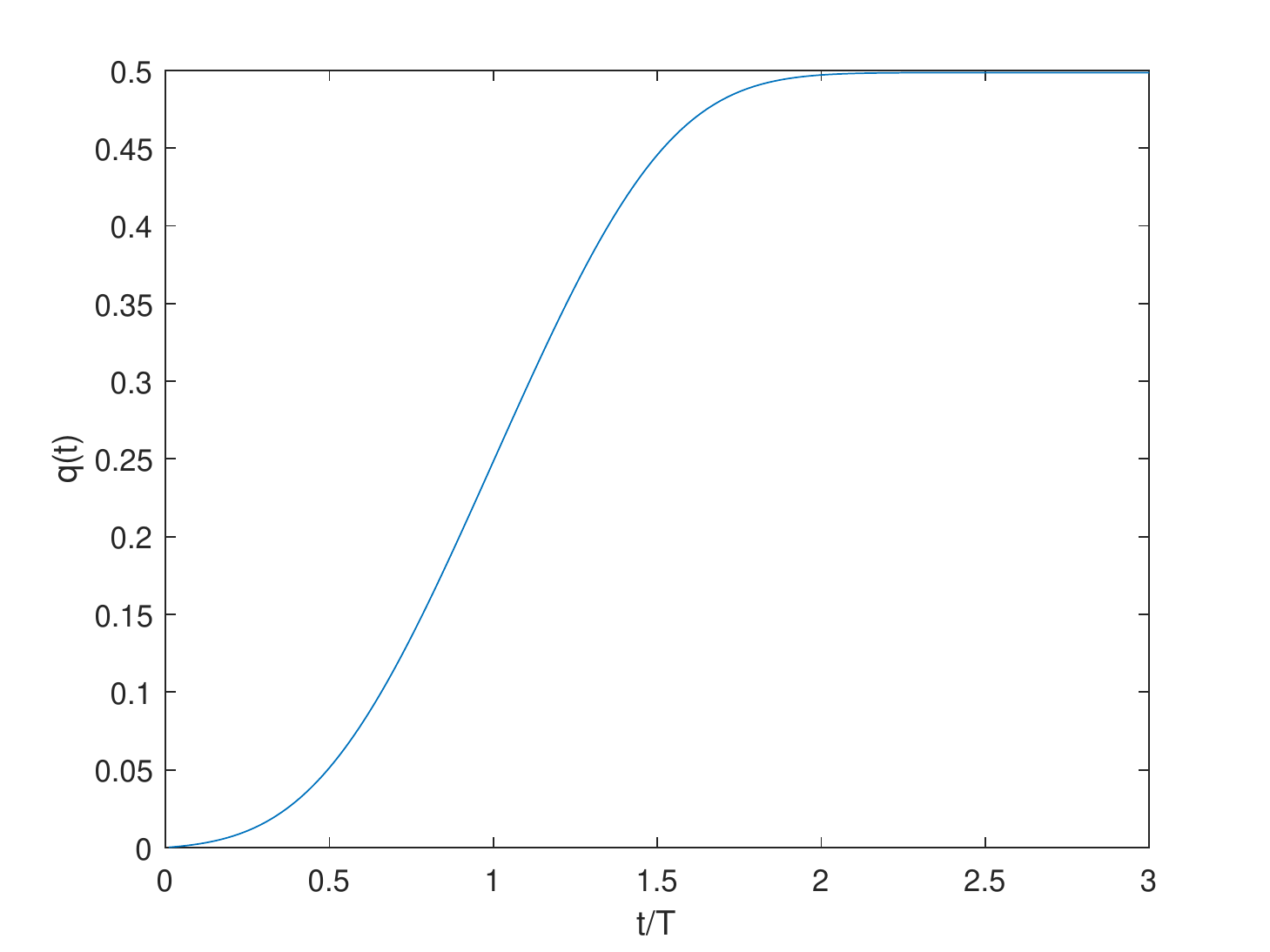 ,width=3.in}
\end{tabular}
\caption{the pulse-shaping function of the GFSK signal with $ BT=0.5 $.}
\label{qt}
\end{figure}

If ${\sf h}=0.5$ (as specified in \cite{au2019bluetooth}), (\ref{eqphin}) becomes
\begin{equation}
\phi_n = \theta_n +\pi I_nq(T),
\label{eqqt}
\end{equation}
where $\theta_n = \frac{\pi}{2}\sum_{m=1}^{n-1} I_m$. Here we set the frequency-time product $BT=0.5$ (also as specified in \cite{au2019bluetooth}). As shown in Fig. \ref{qt}, $q(t) = 0$ for $t\leq0$, and $q(t) \approx 0.5$ for $t\geq2T$.

\begin{figure}[!ht]
\centering
\hspace{-2em}
\begin{tabular}{l}
{\psfig{figure=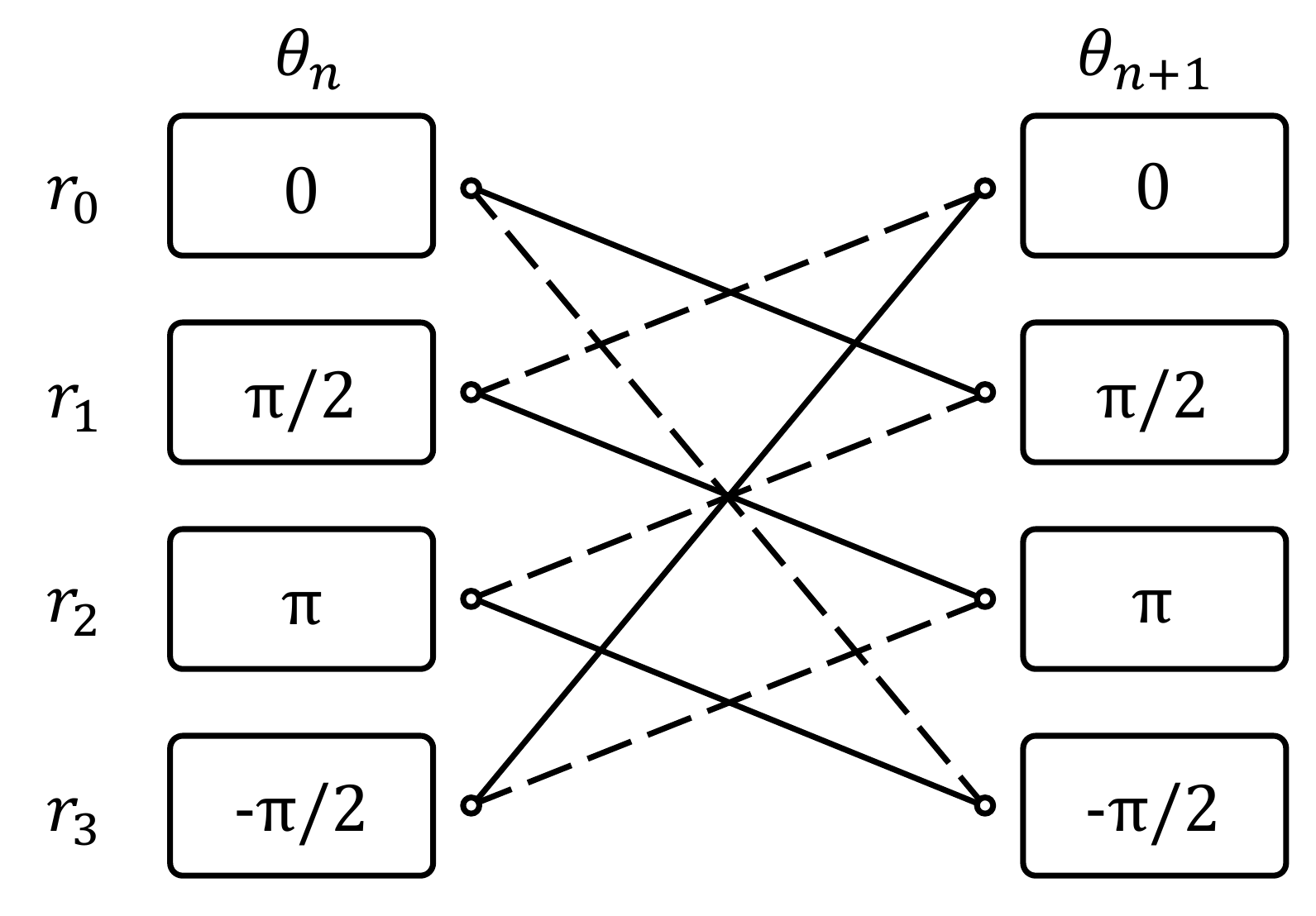 ,width=2.2in}}
\end{tabular}
\caption{The phase transition of the $\theta_n$. The dash line means the input $I_n=-1(b_n = 0)$, solid line means $I_n=1(b_n = 1)$.}
\label{phasetransition}
\end{figure}

We can model the continuous phase modulation as a process of finite state transition based on the phase transition (it is $\theta_n$ not $\phi_n$) as shown in Fig. \ref{phasetransition}. Denote $(r_i,r_j)$ as the state transition from $r_i \in {\mathcal S}$  at time $nT$ to $r_j \in {\mathcal S}$ at time $(n+1)T$ driven by $I_n$. Here $\mathcal {S}$ stands for the set of the states of the trellis.
The dash branches correspond to $I_n=-1$, the solid branches correspond to $I_n = 1$, and the associated signal is
\begin{equation} x_n  = e^{j(\theta_n + \pi I_n q(T))}
= e^{j(\theta_n + \frac{\pi}4 I_n )},\end{equation}
where we have used the fact that $ q(T) = \frac{1}4$.

If the input to the ISI channel (\ref{eq1}) is an GFSK signal, we can combine the GFSK modulation and the multi-path channels into a (larger) trellis diagram.

\subsection{Primer of The Viterbi and BCJR Algorithms}\label{viterbi}
Based on the trellis diagram, the receiver can use the Viterbi algorithm or the BCJR algorithm for optimal symbol detection. We briefly review them to make this paper self-contained.

The Viterbi algorithm is for maximum likelihood (ML) detection. By exploiting the Markovian structure of the finite-memory channel, it computes the likelihood of each branch, i.e., the conditional probability density function (PDF) of the channel output given the inputs
\begin{equation}
p(\mathbf{y}_{1:N}|\mathbf{x}) = \prod_{k=1}^{N} p(y_k|\mathbf{x}_{k-L+1}^{k}),
\label{eq2}
\end{equation}
where $x_k = 0, k<0$.
For a given channel output $\mathbf{y}$, we should have
\begin{equation}
\begin{aligned}
\mathop{\arg\max}\limits_{x_k \in \mathcal{X}} p(\mathbf{y}_{1:N}|{\mathbf{x}})
= \mathop{\arg\min}\limits_{x_k \in \mathcal{X}}-\log p(\mathbf{y}_{1:N}|{\mathbf{x}}).
\end{aligned}
\label{eq3}
\end{equation}
Combining (\ref{eq3}) and (\ref{eq2}), we can rewrite (\ref{eq2}) as:

\begin{equation}
\begin{aligned}
&\mathop{\arg\min}\limits_{x_k \in \mathcal{X}}-\log p(\mathbf{y}_{1:N}|{\mathbf{\emph{x}}}) \\
 =& \mathop{\arg\min}\limits_{x_k \in \mathcal{X}}- \sum_{k=1}^{N} \log  p(y_k|s_k,s_{k+1})
\end{aligned}
\label{eq4}
\end{equation}
where $p(y_k|s_k,s_{k+1})$ is the likelihood of $y_k$ associated with the state transition from $s_k$ to $s_{k+1}$. The Viterbi algorithm can solve (\ref{eq4})  efficiently by searching for the shortest path across the trellis diagram.

The BCJR algorithm \cite{bahl1974optimal} is for maximum \emph{a posteriori} probability (MAP) detection. It computes the \emph{a posteriori} log-likelihood ratio (LLR) 
of bits $\{u_k \}$
\begin{equation}
\begin{aligned}
L(u_k|\mathbf{y})&=\log \frac{p(u_k= 1|\mathbf{y})}{p(u_k= 0|\mathbf{y})}=\log \frac{p(u_k= 1,\mathbf{y})}{p(u_k= 0,\mathbf{y})} \\
&=\log \frac{\sum_{(s_k,s_{k+1})\in \mathcal{S}^{+}}p(s_k,s_{k+1},\mathbf{y})}{\sum_{(s_k,s_{k+1})\in \mathcal{S}^{-}}p(s_k,s_{k+1},\mathbf{y})},
\label{eq6}
\end{aligned}
\end{equation}
where $\mathcal{S}^{+}$, $\mathcal{S}^{-}$ are the set of the ordered pairs $(s_k,s_{k+1})$ corresponding to all states transitions  driven by $u_k = 1$ and $u_k = 0$, respectively.
The sequence $\mathbf{y}$ in $p(s_k,s_{k+1},\mathbf{y})$ can be written as $p(s_k, s_{k+1}, (y_1,\cdots,y_{k-1}), y_k, (y_{k+1}, \cdots, y_N))$. Applying the chain rule for joint probabilities, we can decompose $p(s_k,s_{k+1},\mathbf{y})$ into:
\begin{equation}
\begin{aligned}
p(s_k,s_{k+1},\mathbf{y}) = \alpha_k(s_k)\gamma_k(s_k,s_{k+1})\beta_{k+1}(s_{k+1}),
\label{eq7}
\end{aligned}
\end{equation}
where $\alpha_k(s_k)\triangleq p(s_k,y_{1:k-1})$, $\gamma_k(s_k,s_{k+1})\triangleq p(y_k,s_{k+1}|s_k)$, and $\beta_{k+1}(s_{k+1})\triangleq p(y_{k+1:N}|s_{k+1})$. They can be recursively computed as
\begin{subequations} \label{eq12}
\begin{equation}
\alpha_k(s_k) = \sum_{s_{k-1}} \alpha_{k-1}(s_{k-1})\gamma_{k-1}(s_{k-1},s_{k}),
\label{eq8}
\end{equation}
\begin{equation}
\beta_k(s_k)= \sum_{s_{k+1}} \beta_{k+1}(s_{k+1})\gamma_{k+1}(s_{k},s_{k+1}),
\label{eq9}
\end{equation}
\begin{equation}
\gamma_k(s_{k},s_{k+1})=p(s_{k+1}|s_k)\cdot p(y_k|s_k,s_{k+1}),
\label{eq10}
\end{equation}
\end{subequations}
with the initialization $\alpha_0(s_0) = 1$ and $\beta_N(s_N) = 1$.

Driven by $u_k$, the probability of the state transition is
\begin{equation} p(s_{k+1}|s_k)=\left\{\begin{array}{cl}
p(u_k = 1), & (s_k, s_{k+1}) \in \mathcal{S}^{+}, \\
p(u_k = 0), & (s_k, s_{k+1}) \in \mathcal{S}^{-}, \\
0, & {\rm otherwise}.
\end{array}\right. \label{eq13}
\end{equation}  

\subsection{About the Likelihood Function}
Examining (\ref{eq4}) and (\ref{eq12}), we see that the likelihood $p(y_k|s_k,s_{k+1})$ plays a key role in both Viterbi and BCJR algorithms. Indeed, the Viterbi algorithm depends solely on the likelihood,  while the BCJR also exploits the {\it a priori} information on the state transition probability $p(s_{k+1}|s_k)$,  which is independent of the CSI [cf. (\ref{eq10}) and (\ref{eq13})].  Hence, $p(y_k | s_k, s_{k+1})$ is the only CSI-dependent component for both algorithms. 

The conventional method calculates $p(y_k|s_k,s_{k+1})$ assuming known CSI and that the channel noise is zero-mean Gaussian with variance $\sigma^2$. Hence the likelihood can be computed to be 
\begin{equation}
p(y_k|s_k,s_{k+1}) = \frac{1}{\sqrt{2\pi }\sigma}\exp\left\{{\frac{-|y_k-v_k|^2}{2\sigma^2}}\right\},
\label{eq5}
\end{equation}
where $v_k =\sum_{l=0}^{L-1}h_l x_{k-l}$ is the channel output associated with the  state transition from $s_{k}$ to $s_{k+1}$ (cf. Fig. \ref{Fig2}).

But when the CSI is unknown or the distribution of the noise is unknown owing to non-Gaussian co-channel interference, the model-based likelihood (\ref{eq5}) will be erroneous, causing severe  degradation to the performance of Viterbi or BCJR. To address this issue, we propose { to use an ANN to learn online the likelihood function $p(y_k|s_k,s_{k+1})$ of the trellis diagram based on a pilot sequence, assuming no CSI nor the noise statistics.}


\section{The OLTD-based Viterbi and BCJR} \label{sec:OLTD-based-BCJR}
This section explains how the OLTD-based Viterbi and BCJR algorithms work. Both algorithms consist of three stages as explained in the following. 

\subsection{Stage One: Train ANN to Learn The Trellis Diagram}

\begin{figure}[!ht]
\centering
\begin{tabular}{l}
{\psfig{figure=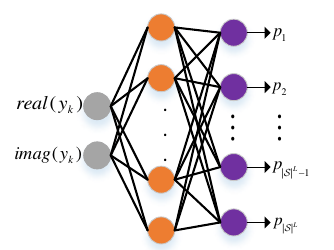, width=2in}}
\end{tabular}
\caption{The one-hidden layer ANN for learning the normalized likelihoods corresponding to the state transitions of the trellis diagram.}
\label{NN}
\end{figure}

As shown in Fig. \ref{NN}, we construct a fully connected and one hidden-layered ANN \cite{lecun2015deep}, whose input is the real and imaginary part of the received sample $y_k$ and the output nodes correspond to all the state transitions in the trellis diagram. For example, we can use a network with 8 output nodes to learn the trellis diagram in Fig. \ref{phasetransition}, which has 8 state transitions. 
The  hidden-layer nodes adopt the Sigmoid activation function, while the output layer employs the \emph{Softmax} activate function. The reason of using the softmax function is to be explained in Section \ref{discussion}.

Given the pilot sequence $\{y_k,k=1,\ldots, P\}$, we know the true transition $s_k\rightarrow s_{k+1}$ corresponding to each $y_k$ and thus label the normalized likelihood $p(y_k|s_k, s_{k+1})$ corresponding to the true transition to be 1 and all the others to be 0; this is the so-called \emph{{one-hot}} encoding. The ANN is optimized according to the minimum cross-entropy criterion. Note that the OLTD method requires neither the CSI nor the noise statistics, which is its main advantage over the model-based method.  

\subsection{Stage Two: Feed the Payload into the ANN}
After the ANN is trained, the payload signals $\{y_i, i=P+1,\ldots, P+Q\}$ are then fed into the network; for each $y_i$ the ANN will yield the normalized likelihoods $p(y_i|s_i, s_{i+1})$ for all the $2|{\mathcal S}|$ state transitions. The numerical examples in Section \ref{V} indicates that the likelihoods yielded by the ANN is sufficiently accurate when it is trained with a pilot with length $P$ no more than a few hundred. 

\subsection{Stage Three: Feed the Likelihoods into the Viterbi or BCJR Algorithm}
Given the likelihoods $p(y_i|s_i, s_{i+1})$'s per received sample of the payload, the Viterbi (or BCJR) algorithm can then be directly applied to obtain the ML (or MAP) detection. 

Taking the BCJR algorithm for example, given the sample $y_k$'s associated  likelihoods $p(y_k|s_k, s_{k+1})$'s, the BCJR algorithm can calculate $\gamma_k$ according to (\ref{eq10}) and further compute the forward recursion (\ref{eq8}) and the backward recursion (\ref{eq9}) to obtain the LLR $L(u_k|\mathbf{y})$ by (\ref{eq7}) and (\ref{eq6}). The combination of the OLTD method and the BCJR algorithm is termed as the OLTD-based BCJR. The OLTD-based Viterbi is even simpler: it just needs to search the most likelihood trellis path according to  (\ref{eq4}). 

\subsection{Discussions}\label{discussion}
Observe from (\ref{eq4}) that to multiply $p(y_k|s_k,s_{k+1})$'s by a common positive factor does not affect the output result of the Viterbi algorithm, neither does it affect the BCJR as can be seen from (\ref{eq6}) and (\ref{eq12}). Hence, instead of learning the original likelihoods, which can be anywhere in $[0,\infty)$, the ANN can simply learn the normalized likelihoods 
\begin{equation} 
\frac{p(y_k|s_k, s_{k+1})}{\sum_{(s_k, s_{k+1}) \in \mathcal{S}^{-} \cup \mathcal{S}^{+}} p(y_k|s_k, s_{k+1})} \in [0, 1], \nonumber
\end{equation}
which is why we can adopt the softmax as the activation function of the output layer even though the actual likelihood value may be out of the range $[0, 1]$. The above insight is the underlying reason why our method is significantly simpler than the state-of-the-art method \cite{shlezinger2020viterbinet}, which has to compute the marginal distribution $p(y_k)$ besides using a deep neural network to learn the conditional probability $p(s_k, s_{k+1}|y_k)$ \cite[Fig. 3]{shlezinger2020viterbinet}.  

We can gain more insights into the OLTD method by drawing its analogy to the classic least square (LS) fitting method. An LS method optimizes the parameters of a signal model to fit the true signal; the OLTD method trains the ANN to approximate the ground truth of the normalized likelihoods, which is a one-hot vector in absence of the channel noise. The LS method uses the LS criteria; the OLTD adopts the minimum cross-entropy criterion. The LS method does not assume knowledge of the noise statistics, neither does the OLTD, which is why the OLTD is robust against non-Gaussian noise as shown in Section \ref{IV}.

Last but not least, our proposed OLTD-based BCJR algorithm can be readily applied for turbo equalization \cite{koetter2004turbo}, which is an important advantage over the BCJRNet \cite{shlezinger2020data} as we will explain in the next.

\section{The OLTD-Based Turbo Equalization for A Coded GFSK System} \label{IV}
\begin{figure}[!ht]
\centering 
\begin{tabular}{l}
{\psfig{figure=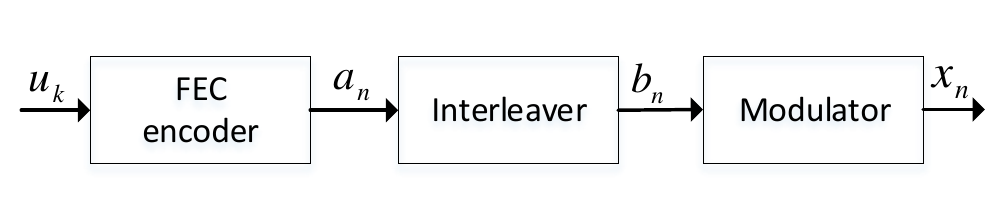 ,width=3.2in}}
\end{tabular}
\caption{A bit-interleaved coded GFSK transmitter, which may be a potential enhancement to the PHY of BLE 5.0.}
\label{bit-interleaver}
\end{figure}

In this section, we apply the OLTD method for turbo equalization, for which a bit interleaver is inserted between the forward error correction (FEC) encoder and the modulator, as shown in Fig. \ref{bit-interleaver}. This section serves for two purposes: i) the OLTD method is shown to be able to be incorporated seamlessly into a turbo equalizer for superb performance; ii) we advocate a potential enhancement to the current BLE standard for significantly better receiver sensitivity, as an interesting byproduct of this study.

Consider that the FEC encoder in Fig. \ref{bit-interleaver} is the convolutional code whose generator polynomials are 
\begin{equation}
\begin{aligned}
G_0(x) &= 1+x+x^2+x^3, \\
G_1(x) &= 1+x^2+x^3,
\end{aligned}
\label{polynomial}
\end{equation}
which is actually the convolutional code adopted in the BLE standard \cite{Bluetooth5}, and the modulation is the GFSK (as explained in Section \ref{coded-GFSK}). The information bits are denoted as $\{u_k\}$, the coded bits $\{a_n\}$, and the interleaved bits $\{b_n\}$, which will be modulated into symbols $\{x_n\}$. 

\subsection{OLTD-Based Turbo Equalization}\label{OLTD-turbo}

\begin{figure}[!ht]
\centering
{\psfig{figure=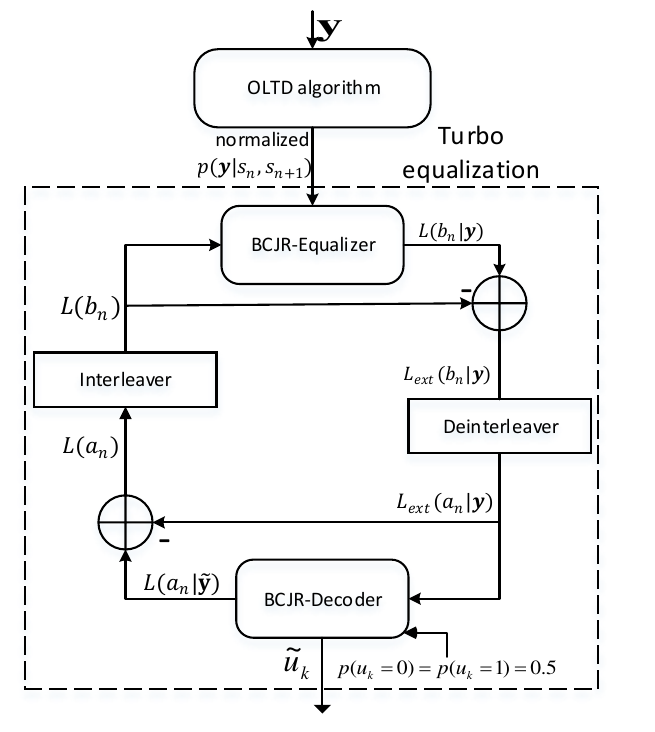 ,width=3.in}}
\caption{Overview of turbo equalization. The OLTD-based turbo equalization differs from the the model-based one only in that the normalized likelihood is obtained using the OLTD algorithm.}
\label{turbo}
\end{figure}

The whole procedure of turbo equalization is as shown in Fig. \ref{turbo}, where BCJR equalization and BCJR decoding are conducted iteratively \footnote{Readers  unfamiliar with turbo equalization may refer to \cite{koetter2004turbo} for an excellent tutorial.}.

On the equalization side, the BCJR algorithm takes the a priori probabilities of $b_n$'s to calculate the state transition probabilities as
\begin{equation}
 p(s_{n+1}|s_n) = 
\begin{cases} p(b_n=0), & \mbox{if $b_n=0$ drives $s_n\rightarrow s_{n+1}$}\\ p(b_n=1), & \mbox{if $b_n = 1$ drives $s_n\rightarrow s_{n+1}$}, 
\end{cases}    \label{eq19a}
\end{equation}
takes the normalized likelihoods from the OLTD algorithm, and then calculate (\ref{eq12}) and (\ref{eq7}) to calculate  $L(b_n|\mathbf{y})$ in (\ref{eq6}). 

Note that the a  posteriori LLR of $\{b_n\}$ can also be computed as
\begin{equation}
\begin{aligned}
L(b_n|\mathbf{y}) &=\log \frac{\sum_{\forall \mathbf{b}:b_n=1}p(\mathbf{b},\mathbf{y})}{\sum_{\forall \mathbf{b}:b_n=0}p(\mathbf{b},\mathbf{y})} \\
&=\log \frac{\sum_{\forall \mathbf{b}:b_n=1}p(\mathbf{y}|\mathbf{b})p(\mathbf{b})}{\sum_{\forall \mathbf{b}:b_n=0}p(\mathbf{y}|\mathbf{b})p(\mathbf{b})}. \\
\label{Lbny}
\end{aligned}
\end{equation}
Since the bit interleaver decorrelates the neighboring coded bits, it holds that $p(\mathbf{b}) = \prod_{n=1}^{N}p(b_n)$; thus, $L(b_n|\mathbf{y})$ can be decomposed into
\begin{equation}
L(b_n|\mathbf{y}) = L_{ext}(b_n|\mathbf{y}) + L(b_n),
\label{14}
\end{equation}
where 
\begin{equation} \label{eqLext}
L_{ext}(b_n|\mathbf{y})=\log \frac{\sum_{\forall \mathbf{b}:b_n=1}p(\mathbf{y}|\mathbf{b})\prod_{i=1:i\neq n}^{N}p(b_i)}{\sum_{\forall \mathbf{b}:b_n=0}p(\mathbf{y}|\mathbf{b})\prod_{i=1:i\neq n}^{N}p(b_i)}
\end{equation}
is the extrinsic information about $b_n$ contained in $\mathbf{y}$, and
\begin{equation}
L(b_n) = \log \frac{p(b_n=1)}{p(b_n=0)}
\label{15}
\end{equation}
is called the intrinsic information. Hence, it follows from (\ref{14}) that 
\begin{equation} L_{ext}(b_n|\mathbf{y}) = L(b_n|\mathbf{y})- L(b_n),\;\; n = 1,2,\ldots,N. \label{eqext}
\end{equation}

On the decoding side, after deinterleaving $L_{ext}(b_n|\mathbf{y})$ into $L_{ext}(a_n|\mathbf{y})$, the BCJR decoder takes the extrinsic information as the ``received signal'', i.e.,  $\tilde{\mathbf{y}} \triangleq L_{ext}(a_n|\mathbf{y})$ to update the a posteriori LLR of the coded bits $L(a_n|\tilde{\mathbf{y}})$ based on the trellis diagram associated with the FEC (\ref{polynomial}). As another input into the BCJR decoder, the uncoded bits $\{u_k,k=1,\ldots, K\}$ are assumed to satisfy $P(u_k=1) = P(u_k=0)=1/2$.

Note from (\ref{14}) that 
we can update the intrinsic LLRs 
\begin{equation}
L(a_n) = L(a_n|\tilde{\mathbf{y}}) - L_{ext}(a_n|\mathbf{y}),
\end{equation}
which can be fed into the BCJR equalizer after being interleaved into $L(b_n)$. Using the relationship [cf. (\ref{15})]
\begin{equation}
p(b_n) = 
\begin{cases} \frac{1}{1+e^{L({b_n})}}, & \mbox{if $b_n=0$,}\\ \frac{e^{L(b_n)}}{1+e^{L(b_n)}}, & \mbox{if $b_n = 1$} \end{cases}
\label{19}
\end{equation}
and (\ref{eq19a}), we can obtain the state-transition probabilities $p(s_{n+1}|s_n)$'s, which are needed for the next round of BCJR equalization.

After a prescribed number of iterations, we calculate $\tilde{\gamma}_k(r_i,r_j)$ as \cite[eq.(22)]{koetter2004turbo}
\begin{equation}
\hspace{-1em}
\begin{array}{l}
\tilde{\gamma}_k(r_i,r_j)= \\
\begin{cases} p(u_k)\cdot p(a_{1,i,j}|\mathbf{y})\cdot p(a_{2,i,j}|\mathbf{y}), & \mbox{if $(s_n,s_{n+1})\in \mathcal{S}^+\cup \mathcal{S}^-$,}\\ 0, & \mbox{if $(s_n,s_{n+1})\notin\mathcal{S}^+\cup \mathcal{S}^-$}, \end{cases}
\end{array}
\label{31}
\end{equation}
for $k=1,2,\cdots,K$, and the decoder will calculate the LLRs of information bits $u_k$, and make a final decision
\begin{equation}
\tilde{u}_k=
\begin{cases} 0, & \mbox{if $L(u_k|\tilde{\mathbf{y}})<0$,}\\ 1, & \mbox{if $L(u_k|\tilde{\mathbf{y}})\geq 0$.} \end{cases}
\label{eq21}
\end{equation}

Hence, the only difference between the OLTD-based turbo equalization and a standard model-based one is how the likelihoods $p(\mathbf{y}|s_n, s_{n+1})$ are obtained, while the whole procedure in the dotted-line box is the same for both methods.

\subsection{Apply BCJRNet to Turbo Equalization?} \label{IV-III}

One may attempt to apply the BCJRNet to turbo equalization, but will come across a major issue as explained in the next.

In the BCJRNet, a neural network is trained to obtain the a posteriori probability $p(s_n,s_{n+1}|y_n)$. Then by Bayes' rule \footnote{To obtain $p(y_n|s_n,s_{n+1})$, the BCJRNet algorithm actually assumes that $p(s_{n+1},s_n) = |{\cal S}|^{-L}$ is a constant \cite[eq. (13)]{shlezinger2020data}.}
\begin{equation}
p(y_n|s_n,s_{n+1}) = \frac{p(s_n,s_{n+1}|y_n)p(y_n)}{p(s_{n+1},s_n)}.
\label{eq24}
\end{equation}

Combining (\ref{eq24}) and (\ref{eq10}), we obtain 
\begin{equation}
    \gamma_n(s_n,s_{n+1}) = \frac{p(s_n,s_{n+1}|y_n)p(y_n)}{p(s_n)}. \label{eq27}
\end{equation}

According to (\ref{eq6}) and (\ref{eq7}), we have
\begin{equation}
\begin{aligned}
&L(b_n|\mathbf{y}) \\
&=\log \frac{\sum_{(s_n,s_{n+1})\in\mathcal{S}^+} \alpha_n(s_n)\gamma_n(s_n,s_{n+1})\beta_{n+1}(s_{n+1})}{\sum_{(s_n,s_{n+1})\in \mathcal{S}^-} \alpha_n(s_n)\gamma_n(s_n,s_{n+1})\beta_{n+1}(s_{n+1})}.
\end{aligned}
\label{eq25a}
\end{equation}

Substituting (\ref{eq27}) into (\ref{eq25a}) yields
\begin{equation}
\begin{aligned}
&L(b_n|\mathbf{y}) \\
&=\log \frac{\sum_{(s_n,s_{n+1})\in\mathcal{S}^+} \alpha_n \cdot\frac{p(s_n,s_{n+1}|y_n)p(y_n)}{p(s_n)}\cdot\beta_{n+1}}{\sum_{(s_n,s_{n+1})\in \mathcal{S}^-} \alpha_n\cdot\frac{p(s_n,s_{n+1}|y_n)p(y_n)}{p(s_n)}\cdot\beta_{n+1}},  \\
&=\log \frac{\sum_{(s_n,s_{n+1})\in\mathcal{S}^+} \alpha_n \cdot\frac{p(s_n,s_{n+1}|y_n)}{p(s_n)}\cdot\beta_{n+1}}{\sum_{(s_n,s_{n+1})\in \mathcal{S}^-} \alpha_n\cdot\frac{p(s_n,s_{n+1}|y_n)}{p(s_n)}\cdot\beta_{n+1}}, 
\end{aligned}
\label{eq25}
\end{equation}

Given the a posteriori probability $p(s_n,s_{n+1}|y_n)$ learned by the BCJRNet, one can update $p(s_n)$ through the forward recursion
\begin{equation}
p(s_n) = \sum_{s_{n-1}\in {\cal S}} p(s_n|s_{n-1}) p(s_{n-1}),
\end{equation} 
where $p(s_n|s_{n-1})$ can be obtained according to (\ref{eq19a}) and (\ref{19}). Due to this recursion, the BCJR-based turbo equalization is cumbersome. 

More important, this method actually does not work as shown by the simulation example in Section \ref{subsec:OOK}. Indeed, the a posteriori probability $p(s_n,s_{n+1}|y_n)$ is not the suitable metric to learn, because 
\begin{equation}
    p(s_n,s_{n+1}|y_n) \propto p(y_n|s_n,s_{n+1})p(s_n,s_{n+1})
\end{equation}
relies on the a priori information $p(s_n,s_{n+1})$; thus, it is impossible to infer $p(s_n,s_{n+1}|y_n)$ from $y_n$ itself unless $p(s_n,s_{n+1})$ is a constant, which is usually untrue. In contrast, the likelihood $p(y_n|s_n,s_{n+1})$ relies solely on the distribution of the channel noise and the CSI, and is independent of the channel coding; thus, it can be learned from $y_n$ itself.

\section{Simulation Results} \label{V}
In this section, we present simulation examples to validate the feasibility and superior performance of the OLTD method applied to the Viterbi algorithm, the BCJR algorithm, and the turbo equalization. 

We adopt a fully-connected neural network with a single hidden layer as shown in Fig. \ref{NN} for the OLTD. The hidden layer has 100 neurons and employs the \emph{Sigmoid} activate function. The number of neurons in the output layer is the same as the number of state transitions in the trellis diagram. The likelihood of each state transition is normalized by the \emph{Softmax} to approximate the ground truth, i.e., the one hot vector. The network is trained using the Adamax optimizer \cite{kingma2015adam} to minimize the cross-entropy based on a pilot sequence. The optimizer divides the pilot into mini-batches of 16 samples and the initial learning rate is set to be 0.01. 
The settings of our training conditions are summarized in Table \ref{tab1}.

\begin{huge}
\begin{table}[!ht]
  \begin{center}
    \caption{The Simulation Setting}
    \setlength{\tabcolsep}{3mm}{
    \begin{tabular}{|c|c|}
      \hline
      NN Toolkit & Keras using Tensorflow backend\\
      \hline
      Training Processor & Inter(R) i7-6700 CPU\\
      \hline
      Training Batch Size & 16  \\
      \hline
      Training Epoch & 200  \\
      \hline
      No. of Hidden Layer& 1 (100 neurons)  \\
      \hline
      Optimizer & Adamax  \\
      \hline
      Activate Function & Sigmoid|Softmax  \\
      \hline
      Loss Function & Cross Entropy\\ 
      \hline
      Pilot Length& 500 in QPSK and OOK, 256 in GFSK cases   \\
      \hline

    \end{tabular}\label{tab1}}
  \end{center}
\end{table}
\end{huge}

Three types of signals are simulated: i)  uncoded Quadrature Phase Shift Keying (QPSK) transmitted over an ISI channel, ii) bit-interleaved On-Off Keying (OOK) in a Poisson channel, and iii) coded GFSK as adopted in the PHY of the BLE -- all can be represented by a trellis diagram. \footnote{The $\rm{Matlab^{TM}}$ codes used for generating the simulation results can be found: https://github.com/JayYang-Fdu/OLTD-code.}.

\subsection{Uncoded QPSK in an ISI Channel with Additive Noise} \label{V-I}

\begin{figure}[!ht]
\centering 
\begin{tabular}{l}
{\psfig{figure=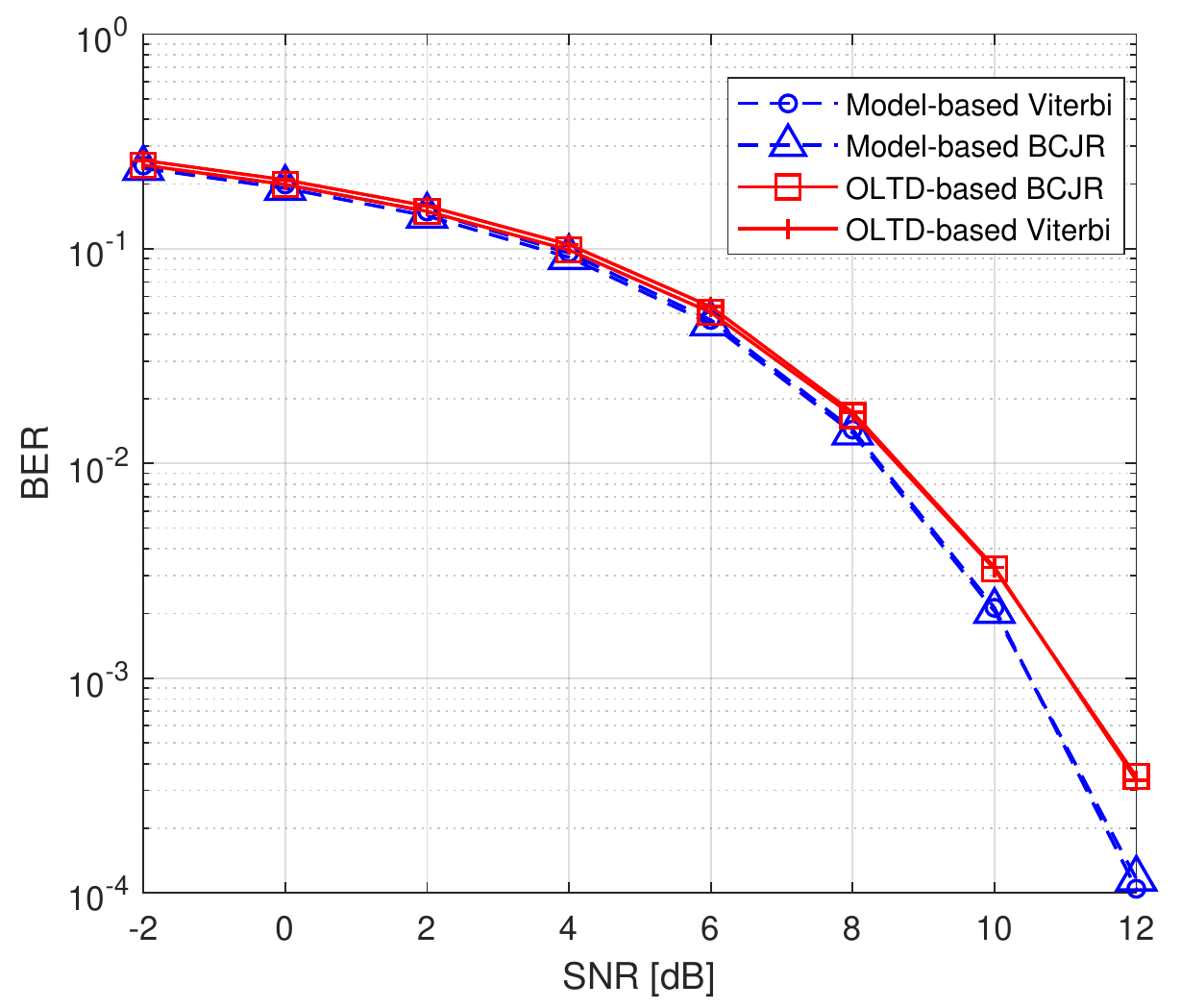,width=3in}}
\end{tabular}
\caption{The BER performance of the model-based and OLTD-based receiver for uncoded QPSK.}
\label{OLTD-BCJR-Viterbi}
\end{figure}

We first simulate an ISI channel as modelled in (\ref{eq1}), where the input signal is uncoded QPSK, the noise is complex-valued Gaussian, and channel coefficients are given by $h_l \triangleq \sqrt {\frac{e^{-\gamma\cdot l}}{\sum^{L-1}_{l=0}e^{-\gamma\cdot l}}}$ for $\gamma > 0$. We set the channel memory length $L = 2$; thus, the trellis diagram is fully-connected with 4 states and 16 branches (state transitions). Fig. \ref{OLTD-BCJR-Viterbi} compares the bit error rate (BER) performance of model-based BCJR/Viterbi algorithm and OLTD-based BCJR/Viterbi algorithm. The model-based method is simulated based on perfect CSI, while the OLTD is trained based on a 500-sample pilot. The simulation results are obtained by averaging over $10,000$ Monte-Carlo simulations, where the channel coefficients are generated with $\gamma$ being draw at random in the range $[0.1,1]$. 
Fig. \ref{OLTD-BCJR-Viterbi} shows that OLTD-based method can achieve performance very close to the model-based benchmark. The Viterbi algorithm performs the same as the BCJR algorithm, since here the bits are generated $0$ or $1$ evenly.

We then simulate the channel noise as complex Cauchy with the PDF
\begin{equation}
f(z_k) = \frac{1}{\pi^2}\cdot \frac{\lambda^2}{(\Re\{ z_k\}^2+\lambda^2)\cdot(\Im\{z_k\}^2+\lambda^2)},
\end{equation}
where 
$\Re\{\cdot\}$ and $\Im\{\cdot\}$ stand for the real and imaginary parts, respectively. The Cauchy's PDF has much heavier tails than the Gaussian's.

\begin{figure}[!ht]
\centering 
\begin{tabular}{c}
{\psfig{figure=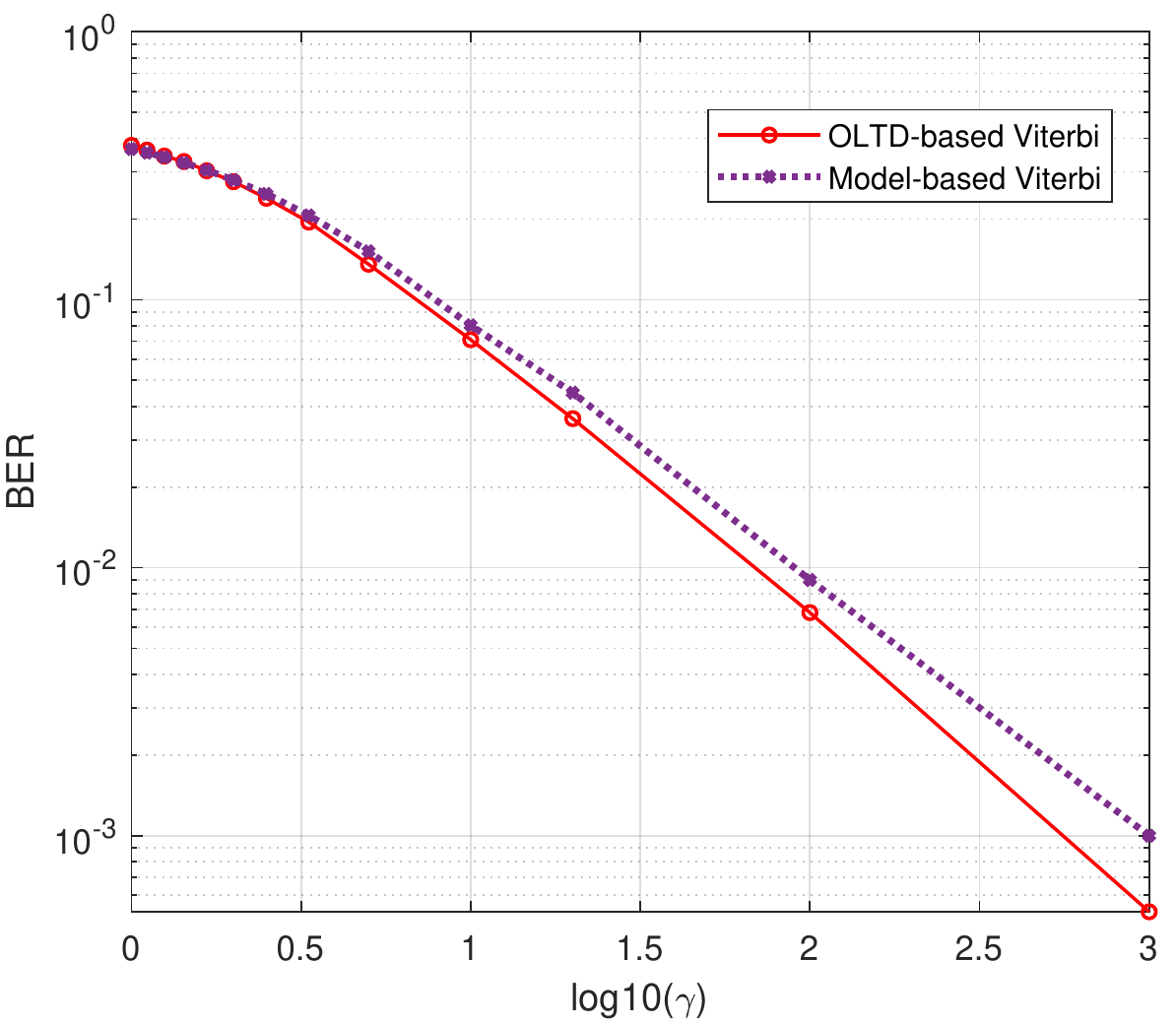 ,width=3in}}
\end{tabular}
\caption{The BER performance of model-based Viterbi receiver OLTD-Viterbi receiver.}
\label{cauchy}
\end{figure}


In the Cauchy noise case, the SNR is undefined since the variance of Cauchy is unbounded. We simulate the BER performance by changing the value of $\gamma$.  Fig. \ref{cauchy} shows that the OLTD-BCJR outperforms the model-based BCJR algorithm in this non-Gaussian case, which is not surprising since the OLTD requires no \emph{a priori} knowledge on the statistics of the noise.

\begin{figure}[!ht]
\centering 
\begin{tabular}{l}
{\psfig{figure=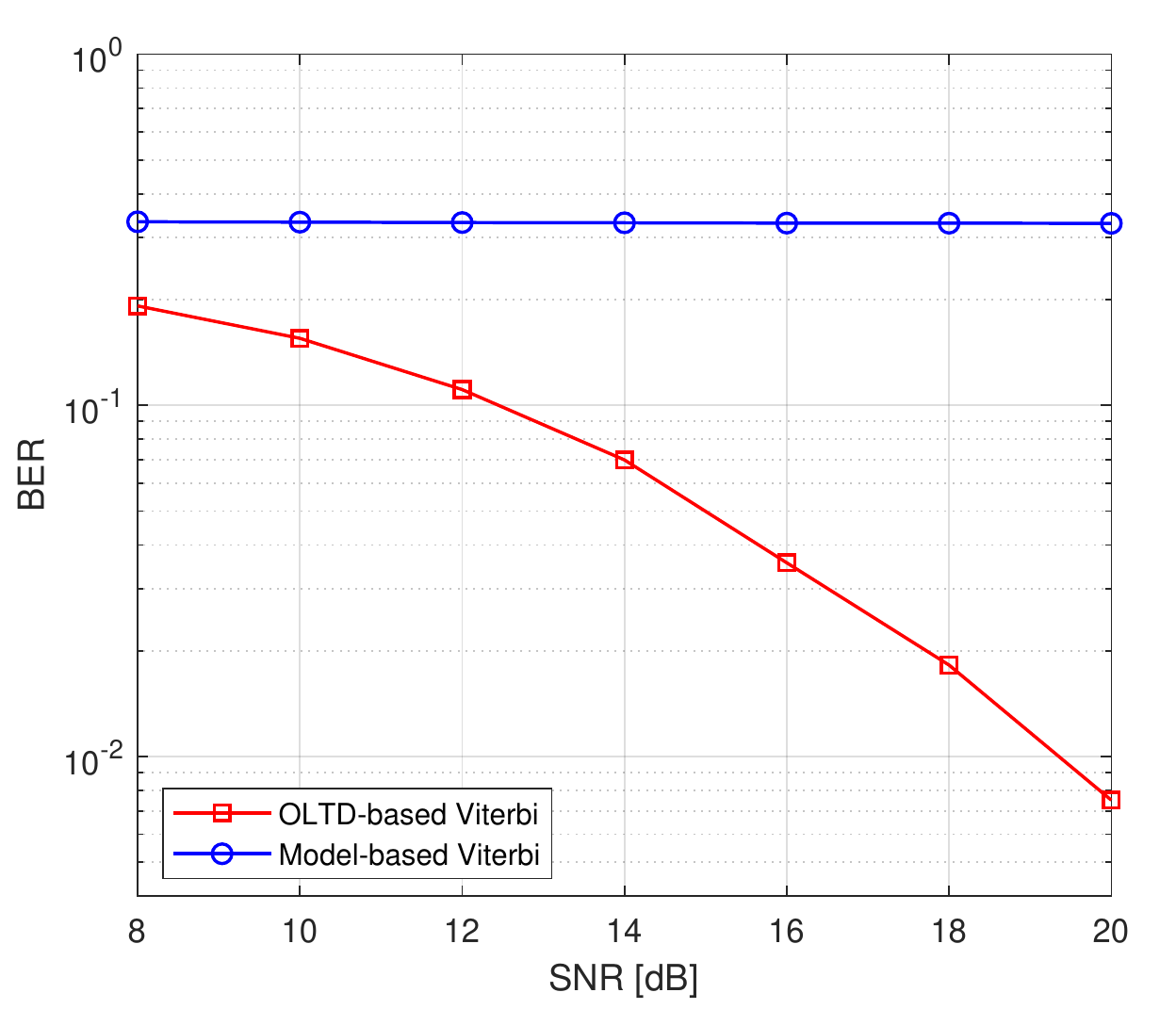,width=2.8in}}
\end{tabular}
\caption{The BER performance of the OLTD-based Viterbi compared to the model-based method with co-channel interference.}
\label{uncoded_qpsk}
\end{figure}

We also simulate the ISI channel scenario where the QPSK signal is interfered by a random  4-ary Pulse Amplitude Modulation (4-PAM) source. The received power of the interference is the same as the QPSK signal, i.e., the signal-to-interference ratio (SIR) is $0$ dB. As shown in Fig. \ref{uncoded_qpsk}, the model-based Viterbi method based on the assumption of Gaussian noise fails. The striking advantage of the OLTD-based approach indicates that the neural network somehow learned the ``structure'' of the non-Gaussian interference and hence suppressed it effectively.


\begin{figure}[!ht]
\centering 
\begin{tabular}{c}
{\psfig{figure=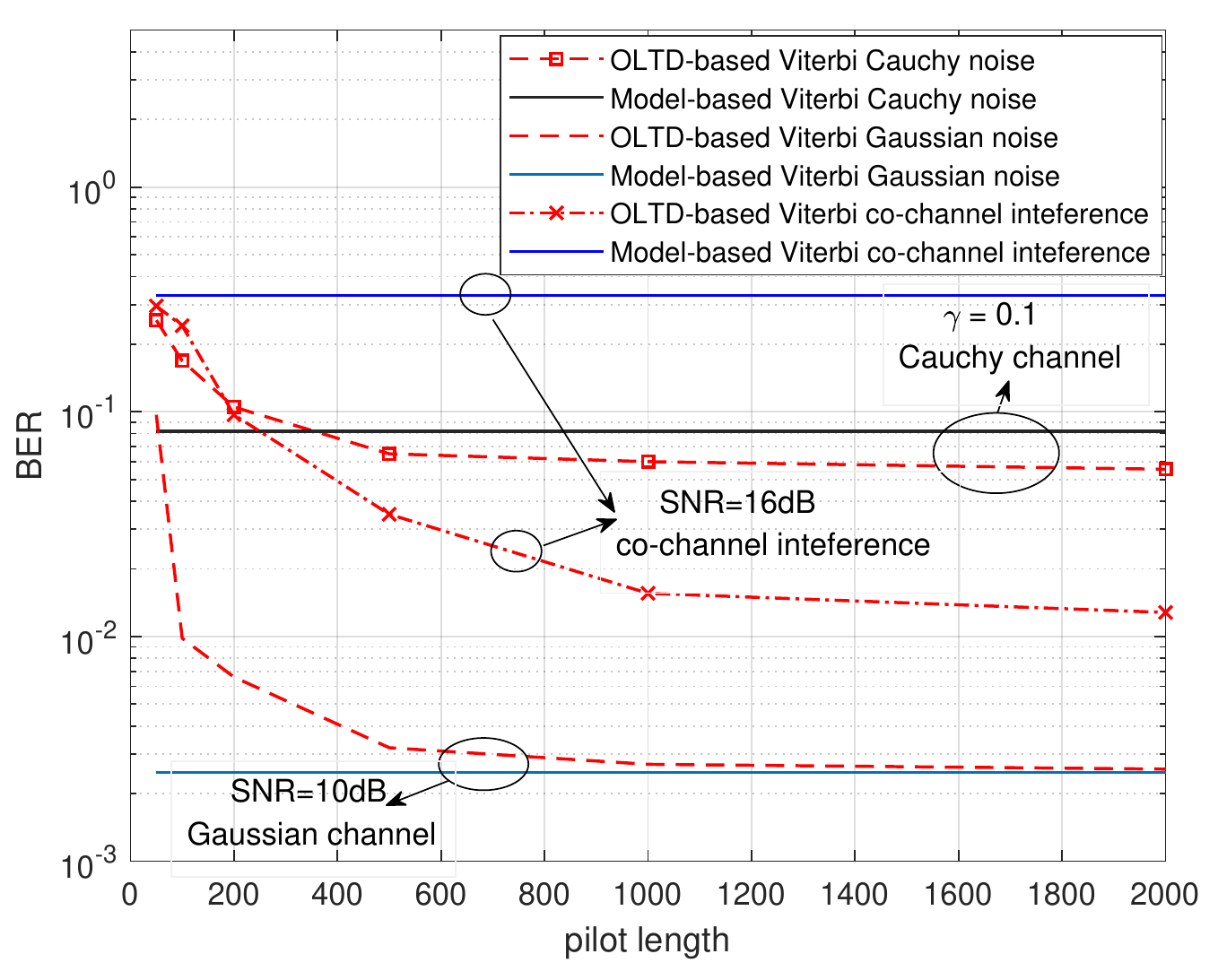 ,width=2.8in}}
\end{tabular}
\caption{The BER performance versus the length of pilots in the SNR value 10dB.}
\label{pilotsvsBER}
\end{figure}

To demonstrate the influence of training pilot length on reception performance, we simulate the BER performance of the OLTD-based approach when trained using the pilots length of $\{0.5, 1, 2, 5, 10, 20\}\times 10^{2}$ as shown in Fig. \ref{pilotsvsBER}. We set  SNR = 10 dB under Gaussian channel, and the performance of the OLTD-based method, as the training pilots length increases, gradually approaches that of the model-based method under perfect CSI. A pilot of length a few hundred is sufficient for all the three cases.  Fig. \ref{pilotsvsBER} illustrates from a different perspective that the OLTD-based method can outperform the model-based method in the two cases of non-Gaussian noise.

\subsection{Bit-interleaved Coded OOK in a Poisson Channel} \label{subsec:OOK}

In addition to the additive noise channel, we also simulate the Poisson channel as previously considered in \cite{shlezinger2020data}. But here we consider a bit-interleaved system as shown in Fig. \ref{bit-interleaver}, where the information bits are FEC encoded, bit-interleaved, and then modulated by  OOK before being transmitted over a Poisson channel. The channel output $y_k$ is of Poisson distribution, i.e.,
\begin{equation}
p(y_k|s_k,s_{k+1} ) =  \frac{(v +1)^{y_k}}{y_k!}e^{-(v +1)},\\ 
 y_k=0,1,2,\ldots
\label{r}
\end{equation}
where $v=\sum_{l=0}^{L-1}h_lx_{k-l}$ with $x_{k-l} \in \{0,1\}$ and $h_l \triangleq \sqrt {\frac{e^{-\gamma\cdot l}}{\sum^{L-1}_{l=0}e^{-\gamma\cdot l}}}$ for $\gamma$ being draw at random in the range$[0.1,1]$. 

\begin{figure}[!ht]
\centering 
\begin{tabular}{l}
{\psfig{figure=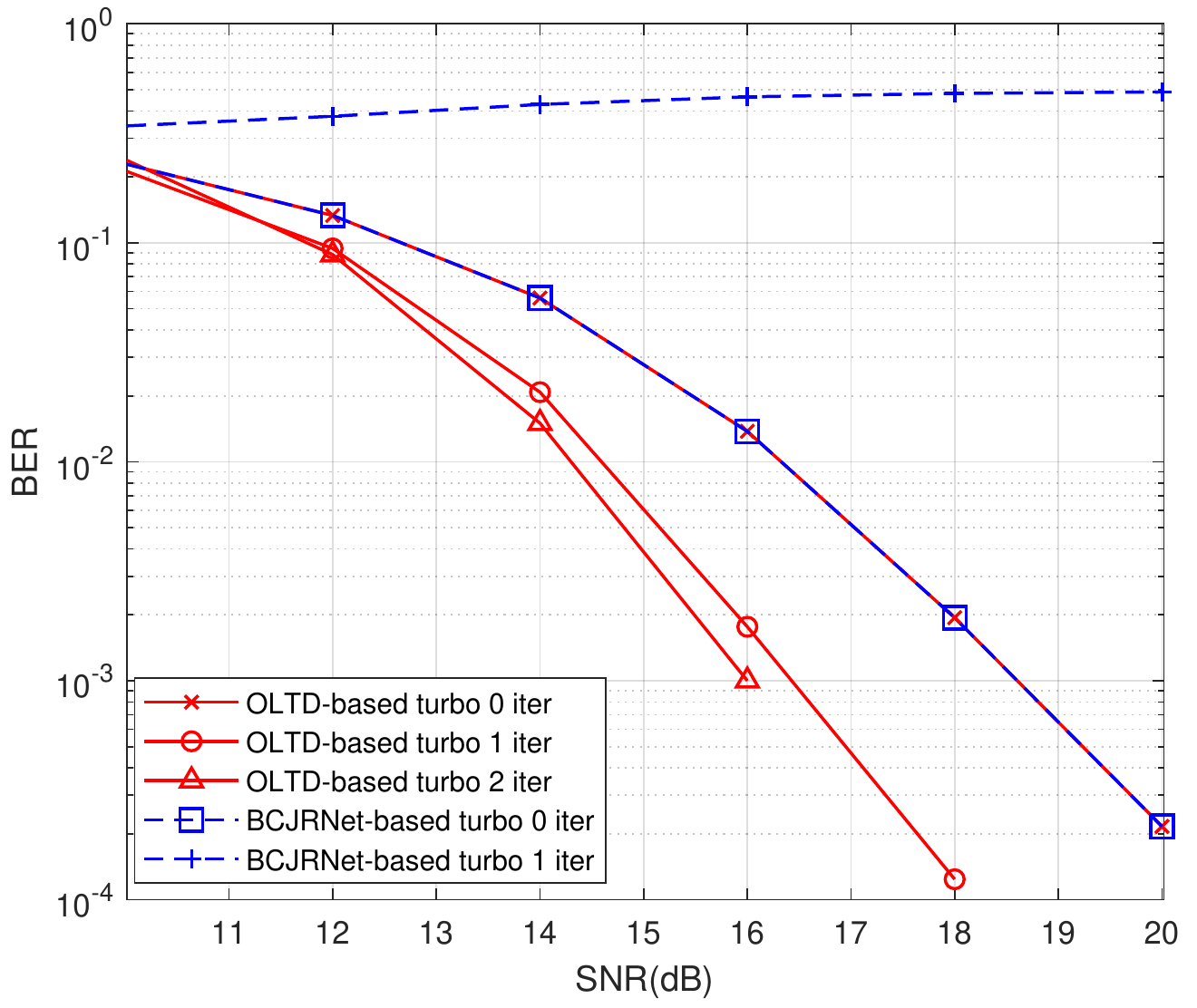,width=2.8in}}
\end{tabular}
\caption{The BER performance of the OLTD-based turbo equalization compared to the BCJRNet-based turbo in Poisson channel.}
\label{turbo_poisson}
\end{figure}

For $L=2$, the Poisson channel can be represented by a fully-connected two-state trellis diagram with 4 branches. Hence, the OLTD uses a neural network with 4 outputs. Fig. \ref{turbo_poisson} shows that using no iterations the OLTD-based method and the BCJRNet have identical performance. The BCJRNet applied for turbo equalization, however, leads to failed decoding as explained in Section \ref{IV-III}. In contrast, the OLTD-based turbo equalization can achieve significantly improved performance. 

\subsection{A BLE System and Its Enhancement}

\begin{figure}[!ht]
\centering 
\begin{tabular}{l}
{\psfig{figure=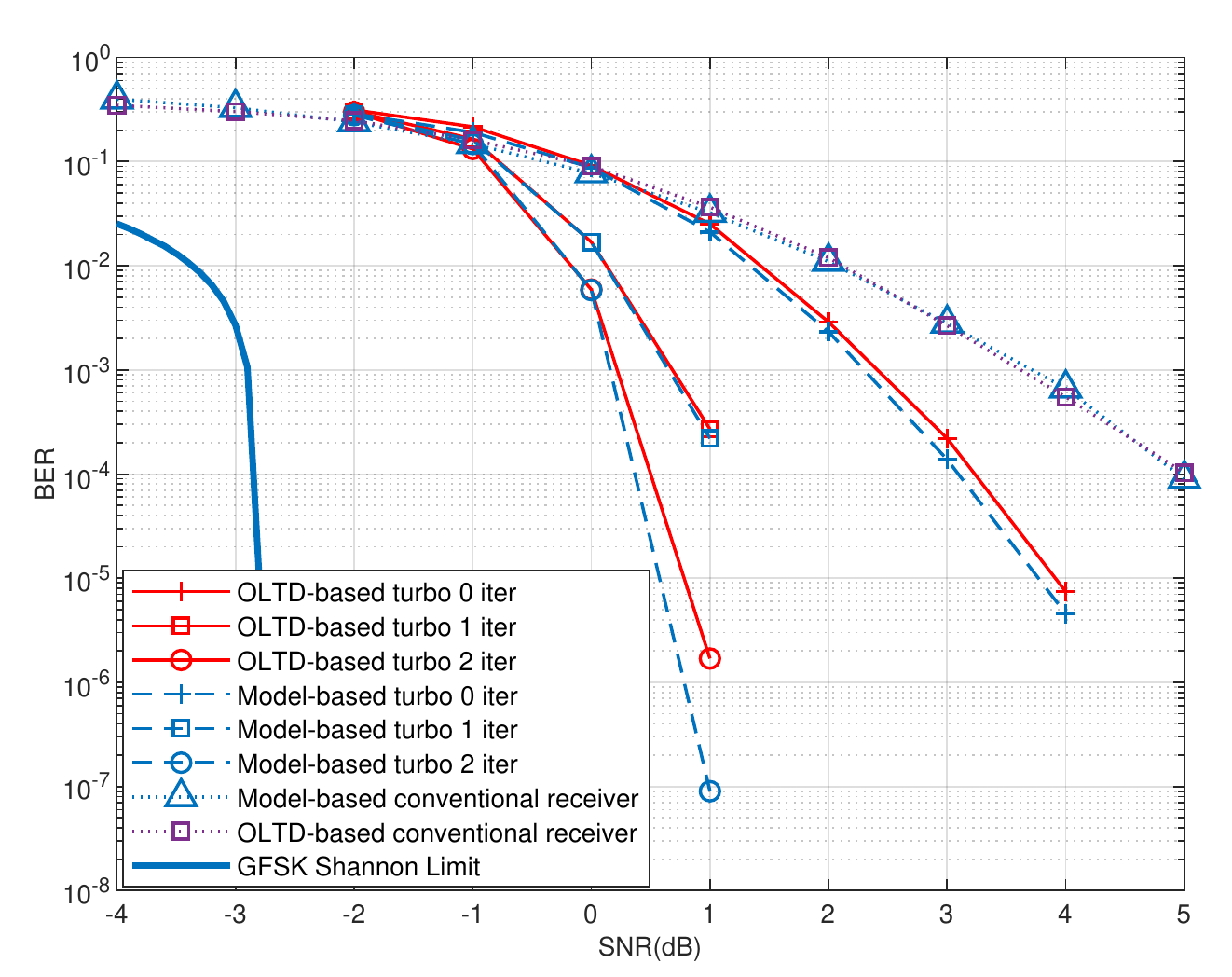,width=2.8in}}
\end{tabular}
\caption{The performances of the model-based and OLTD-based BLE receiver in an AWGN channel versus the performance of the enhanced BLE system with bit-interleaving.}
\label{OLTD_turbo256}
\end{figure}

We simulate in an AWGN channel a BLE system with bit rate 500 Kbps with a pilot of length  256 as specified in the protocol \cite{Bluetooth5}. A model-based receiver consists of a BCJR demodulator and a Viterbi decoder. It assumes perfect CSI. The OLTD-based receiver differs from the model-based one only in that the likelihoods $p(y_k|s_k,s_{k+1})$ [cf. (\ref{eq10})] used in the  BCJR algorithm is produced by a neural network trained by the 256-sample pilot sequence. According to  Fig. \ref{phasetransition}, the GFSK modulation can be represented by a trellis diagram with 8 state transitions; thus, the neural network for the OLTD has 8 neurons in the output layer. The BER is averaged over $10,000$ Monte Carlo trails of the channel coefficients $h(\theta) = e^{j\theta}$, with $\theta$ being drawn at random in the range $[0, 2\pi]$. In Fig. \ref{OLTD_turbo256}, the dot dash line with marker $\triangle$ corresponds to the model-based method, while the dot dash line with marker $\square$ is the OLTD-based one. They essentially overlap, which suggests that the neural network assisted receiver can be practically feasible at least performance-wise. 

We also present Shannon limit of the BER performance of the GFSK signal obtained using the numerical method in \cite{2014Fundamental, 1661831}. The large gap between Shannon limit and the achieved performance of the BLE system motivated us to consider introducing a bit interleaver at the transmitter between encoder and GFSK modulator (cf. Fig. \ref{bit-interleaver}). Given the bit-interleaver, the receiver can apply the turbo equalization. Fig. \ref{OLTD_turbo256} also illustrates the BER performance of the turbo receiver with no iteration, and with 1 and 2 iterations. The three dash lines show the model-based turbo equalization under the perfect CSI and the other three sold lines with markers correspond to the OLTD-based turbo algorithm with pilot length of 256.
It can be seen that introducing the bit-interleaver and using the turbo equalization in the receiver can yield $4\sim 5$dB gain compared with the conventional receiver. Hence, to introduce bit-interleaving may be an interesting enhancement to the existing BLE protocol for significantly enhanced performance, which can make it more competitive for IoT communications.

\begin{figure}[!ht]
\centering 
\begin{tabular}{l}
{\psfig{figure=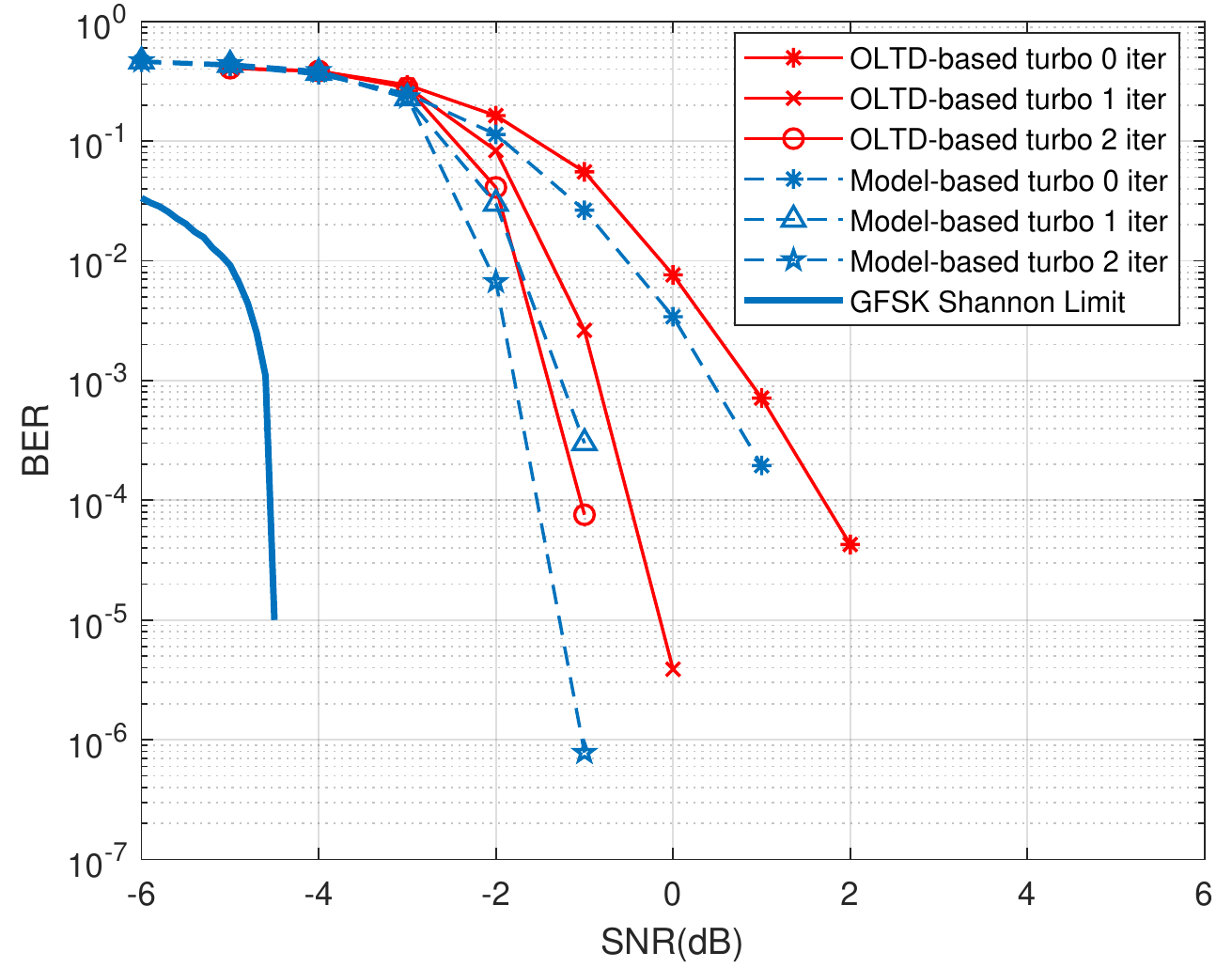 ,width=2.8in}}
\end{tabular}
\caption{The BER performance of the OLTD-based turbo equalization in ISI channel with channel length $L=2$.}
\label{ISI}
\end{figure}

In the last example, we consider the BLE system in a ISI channel with memory $L=2$. The ISI channel is described as $h_0\delta(n) + h_1\delta(n-1)$, where the channel coefficients are normalized. The combination of the GFSK modulation and the ISI channel can be modeled by a trellis diagram with 16 states. Then we can also apply turbo equalization based on the OLTD method except that here the output layer of the neural network has $16$ neurons. The BER performance of the model-based turbo receiver with known perfect CSI and the OLTD-based turbo receiver in the ISI channel is compared in Fig. \ref{ISI}. It can be seen that the OLTD-based turbo equalization algorithm trained based on a same 256-sample  pilot sequence has about 0.5dB loss compared with the model-based method with perfect CSI.

\section{Conclusions} \label{conclusion}
This paper introduced a method named online learning of trellis diagram (OLTD), which uses a single hidden-layer artificial neural network (ANN) to learn the likelihoods of the received samples under different state transitions. It can be applied to replace only the channel state-dependent part of the Viterbi algorithm and the BCJR algorithm. We applied the OLTD-based Viterbi/BCJR algorithms to a coded QPSK/GFSK system, and the simulation results show that using a pilot sequence of length only a few hundred samples the OLTD based methods can perform similarly to their model-based counterpart given perfect channel state information (CSI) and Gaussian noise. 
In contrast to  the model-based approaches, the OLTD-based approach assumes neither CSI nor statistics of the noise, which makes it robust against non-Gaussian interferences. In contrast to the state-of-the-art machine learning assisted methods, such as the BCJRNet and ViterbiNet, the proposed method does not assume the a priori probabilities of the coded bits, which makes it readily applicable to turbo equalization. The OLTD-based algorithms can be applied to the standard Bluetooth system and the enhanced one with bit-interleaving, because they require only some pilot of moderate length. Introducing  bit-interleaving can be a beneficial enhancement to the BLE standard as an interesting by-product of this study. 

\bibliographystyle{ieeetr}
\bibliography{all}

\end{document}